\documentclass[referee,pdflatex,sn-mathphys]{sn-jnl}

\usepackage{sah_shortcuts}
\usepackage{adjustbox}

\jyear{2023}

\begin{document}

\title[Extreme Indian rainfall under El Ni\~no]{More extreme Indian monsoon daily rainfall in El Ni\~no summers}


\author*[1]{\fnm{Spencer A.} \sur{Hill}}\email{shill1@ccny.cuny.edu}
\author[2]{\fnm{Destiny Zamir} \sur{Meyers}}\email{d.z.meyers@columbia.edu}
\author[3,4]{\fnm{Adam H.} \sur{Sobel}}\email{ahs129@columbia.edu}
\author[3]{\fnm{Michela} \sur{Biasutti}}\email{biasutti@ldeo.columbia.edu}
\author[3]{\fnm{Mark A.} \sur{Cane}}\email{mcane@ldeo.columbia.edu}
\author[4]{\fnm{Michael K.} \sur{Tippett}}\email{mkt14@columbia.edu}
\author[5]{\fnm{Fiaz} \sur{Ahmed}}\email{fiaz@ucla.edu}

\affil*[1]{\orgdiv{Department of Earth and Atmospheric Sciences}, \orgname{City College of New York}, \orgaddress{\street{160 Convent Avenue}, \city{New York}, \postcode{10031}, \state{New York}, \country{United States}}}

\affil[2]{\orgdiv{Department of Civil Engineering \& Engineering Mechanics}, \orgname{Columbia University}, \orgaddress{\street{500 W. 120th Street \#610}, \city{New York}, \postcode{10027}, \state{New York}, \country{United States}}}

\affil[3]{\orgdiv{Lamont-Doherty Earth Observatory}, \orgname{Columbia University}, \orgaddress{\street{61 Route 9W}, \city{Palisades}, \postcode{10964}, \state{New York}, \country{United States}}}

\affil[4]{\orgdiv{Department of Applied Physics and Applied Mathematics}, \orgname{Columbia University}, \orgaddress{\street{500 W. 120th Street \#200}, \city{New York}, \postcode{10027}, \state{New York}, \country{United States}}}

\affil[5]{\orgdiv{Department of Atmospheric and Oceanic Sciences}, \orgname{UCLA}, \orgaddress{ \city{Los Angeles}, \postcode{90095}, \state{California}, \country{United States}}}


\abstract{%
Extreme rainfall in the Indian summer monsoon can be destructive and deadly \cite{noauthor_483_2018}.  Although \elnino/ events in the equatorial Pacific make dry days and whole summers more likely throughout India \cite{walker_meteorological_1910,pant_aspects_1981,rasmusson_relationship_1983}, their influence on daily extremes is not well established.  Despite this summer-mean drying effect, we show using observational data spanning 1901-2020 that \elnino/ increases extreme rainfall likelihoods within monsoonal India, especially in the the summer's core rainy areas of central-eastern India and the narrow southwestern coastal band.  Conversely, extremes are broadly suppressed in the drier southeast and far northwest, and more moderate accumulations are inhibited throughout the domain.  These rainfall signals appear driven by corresponding ones in convective buoyancy, provided both the undilute instability of near-surface air and its dilution by mixing with drier air above are accounted for \cite{ahmed_deep_2020}.  When the summer ENSO state is predicted from a seasonal forecast ensemble initialized in May, the extreme rainfall patterns broadly persist, suggesting the potential for skillful seasonal forecasts.  The framework of analyzing the full distributions of rainfall and convective buoyancy could be usefully applied to hourly extremes, other tropical regions under ENSO, other variability modes, and to trends in extreme rainfall under climate change.
}



\maketitle

The influence of the \enso/ (ENSO) on total summer rainfall in India has been recognized since the early 20th century \cite{walker_meteorological_1910,pant_aspects_1981,rasmusson_relationship_1983,gadgil_indian_2003}.  Anomalously warm equatorial Pacific sea surface temperatures (SSTs) in \elnino/ summers generate anomalous season-mean ascent locally but compensating descent over much of the rest of the tropics including India.

One might expect summer-mean subsidence to inhibit moist convection of all intensities, extremes included, reducing both the mean and variance of daily rainfall.  Indeed, recent studies \cite{moron_spatial_2017,hill_all-india_2022} establish that \elnino/ reduces the average accumulation on rainy days---the rainfall intensity---within southeastern India and in Rajasthan state in the northwest.  These are the summer monsoon's climatologically dry areas, and within them during \elnino/ summers it tends to rain even less overall, less often, and less intensely  \cite{moron_spatial_2017,hill_all-india_2022}.  

But the opposite rainfall intensity signal emerges in the summer monsoon's climatologically rainy areas \cite{moron_spatial_2017,hill_all-india_2022}: the broad Central Monsoon Zone in central India and the narrow southwestern coastal band of peninsular India.  Within them during \elnino/ summers it tends to rain less overall and less often---but more intensely when the rain comes.

Whether these signals in rainfall intensity stem from destructive extremes \cite{suhas_influence_2023} or beneficial moderate events \cite{gadgil_indian_2006}---and mechanistically what causes them---remain unanswered.  Observational studies have investigated ENSO's influence on daily or hourly rainfall extremes in other tropical domains \cite{curtis_precipitation_2007,allan_atmospheric_2008,revadekar_nino-southern_2008,grimm_enso_2009,sun_global_2015,saunders_spatial_2017,li_global_2020,vashisht_modulation_2021} but not the Indian summer monsoon.

\section*{ENSO influences on extreme daily rainfall}
\label{sec:results}

\begin{figure*}
\centering
\adjustimage{valign=c,width=0.33\textwidth}{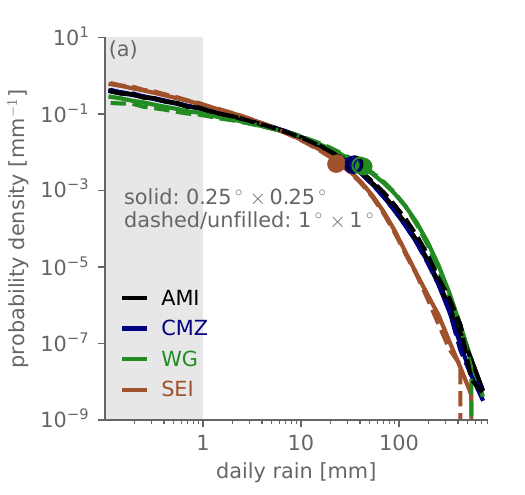}
\adjustimage{valign=c,width=0.66\textwidth}{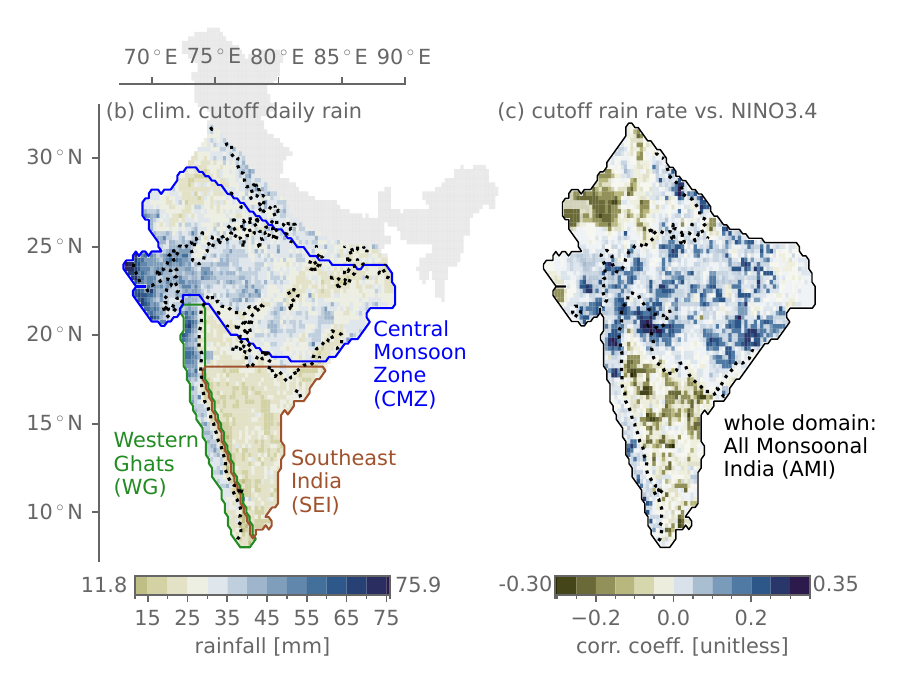}\\
\caption{
  (a) Daily rainfall empirical probability distributions (Methods), in log-log space, combining all June-July-August-September (JJAS) days 1901-2020 and all gridpoints within all of monsoonal India (black), the Central Monsoon Zone (blue), the Western Ghats coastal band (green), or Southeast India (brown).  Overlaid circles indicate the climatological cutoff.
  (b) Local climatological cutoff daily rainfall (mm) across all JJAS days 1901-2020 according to the colorbar.
  The portions of northern and northeast India shaded grey are excluded from all analyses (Methods).
  The dotted black contour denotes the median across gridpoints of the climatological 99th percentile daily rain rate (63.1~\mmday/).
  (c) Pearson correlation coefficient between the JJAS NINO3.4 value and each gridpoint's cutoff rain rate.  The dotted black contour denotes the JJAS climatological rainfall median (5.9~\mmday/) across gridpoints.  Semi-transparent shading indicates statistical insignificance at the 95\% confidence level based on a block-bootstrap method.  All timeseries are linearly detrended over 1901-2020 prior to correlations being computed.
}
\label{fig:corr-maps}
\end{figure*}

Our primary metric of extreme daily rainfall is the cutoff accumulation \cite{neelin_global_2017,neelin_precipitation_2022}, computed as the ratio of the variance to the mean of daily rainfall across summer days (Methods).  This metric is motivated by the fact that daily rainfall reliably follows a gamma distribution \cite{martinez-villalobos_why_2019}: probabilities follow a shallow power-law below the cutoff before dropping off exponentially above it (Fig.~\ref{fig:corr-maps}a).  Increasing the cutoff, which in the exact gamma limit is identical to the scale parameter, therefore makes all extreme rain rates more likely.  Our primary dataset is the Indian Meteorological Department daily 0.25\(\times\)0.25\degr{} gridded rainfall product spanning 1901-2020 derived from a dense, temporally varying in situ rain gauge network \citep{pai_development_2014}.  We use all 122 days within each June-July-August-September (JJAS) summer season and all gridpoints within the ``monsoonal India'' domain \cite{gadgil_new_2019} (Methods).  As shown below, key results are robust across alternative metrics of extremes and to another daily rainfall dataset.

The climatological cutoff spans from 11.8 to 75.9~mm across all monsoonal India gridpoints (Fig.~\ref{fig:corr-maps}b), with a spatial structure that largely tracks that of summer-mean rainfall: high values within the southwestern coastal band, a sharp gradient moving east across the Western Ghats mountain range to much smaller values in the southeast, intermediate values north thereof in most of the Central Monsoon Zone, and low values again in the far northwest.  This pattern emerges in many properties of summer monsoon rainfall from daily to interannual timescales \cite{krishnamurthy_intraseasonal_2000,moron_spatial_2017,hill_all-india_2022} (Fig.~\ref{fig:corr-maps}a; overlaid contour in Fig.~\ref{fig:corr-maps}b; Extended Data Fig.~\ref{fig:clim-and-stdevs}).  The cutoff and other metrics of extremes are large compared to the mean in Gujarat (along the Arabian Sea coast within the Central Monsoon Zone), which is rather dry most summers but in others receives synoptically driven extreme accumulations \cite{hunt_relationship_2019} (Extended Data Fig.~\ref{fig:3points}).

Based on lag-zero correlations of each gridpoint's JJAS cutoff with the standard NINO3.4 index of ENSO (Methods), extreme daily rainfall becomes more likely over much of the southwestern coastal band and central India as NINO3.4 increases but less likely over much of the southeast and far northwest (Fig.~\ref{fig:corr-maps}c).  This pattern resembles the climatological summer-mean rainfall (overlaid contour in Fig.~\ref{fig:corr-maps}c; Extended Data Fig.~\ref{fig:clim-and-stdevs}) and the established \cite{moron_spatial_2017,hill_all-india_2022} ENSO-driven patterns in rainfall intensity.  As one quantification of this resemblance, of the 581 gridpoints with statistically significant positive correlations, 75\% occur in gridpoints with climatological summer-mean rainfall above the median across gridpoints, and of the 384 gridpoints with significant negative correlations, 99\% occur in gridpoints below it.  As another, linear regressions of each gridpoint's NINO3.4-cutoff correlation \vs/ its summer-mean climatological rainfall are significantly positive for the whole domain and within each subregion (Extended Data Fig.~\ref{fig:mean-vs-n34corr}).

\begin{figure*}
\centering
\includegraphics[width=0.33\textwidth]{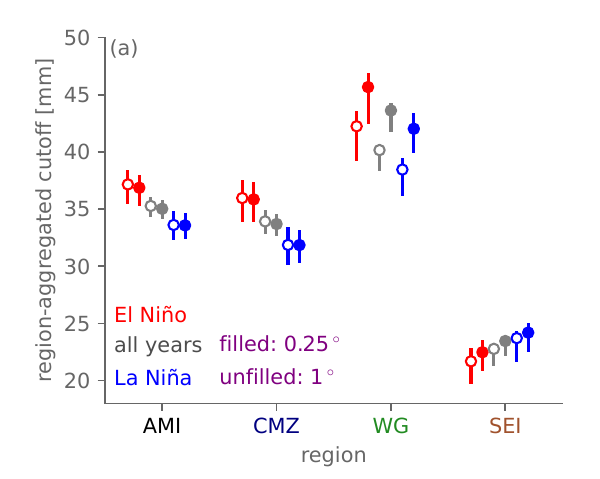}
\includegraphics[width=0.66\textwidth]{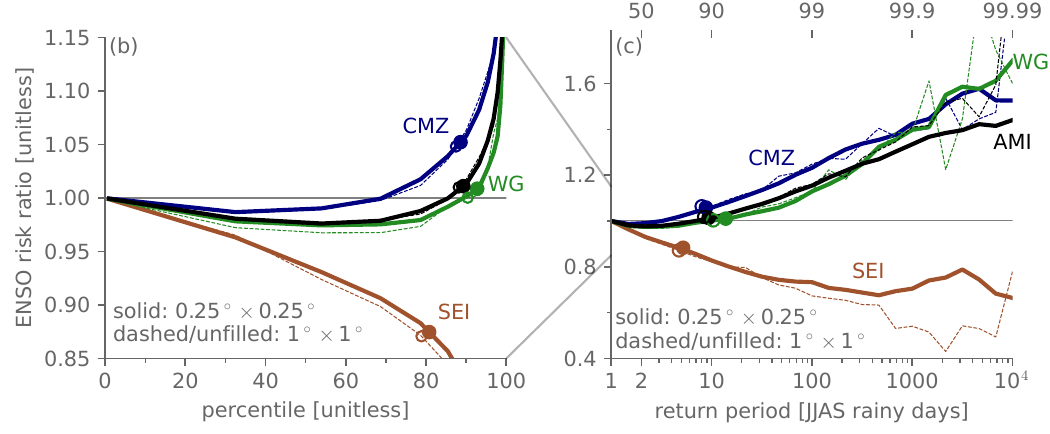}
\caption{
  (a) Composite regional cutoffs (in~mm) in circles.
  Bars are 95\% confidence intervals based on yearly cutoff values, and squares are the estimates from gamma fits (Methods).
  (b) ENSO risk ratios: restricting to rainy days (\(>1\) mm), the probability that rainfall exceeds a given rain rate in the \elnino/ composite divided by that in the \lanina/ composite.  The horizontal axis is the percentile daily rain rate across all rainy days in all years and points for that region.  Overlaid circles show each region's climatological cutoff.
  (c) Same as panel c, but with logarithmically spaced horizontal axis and wider vertical axis to better discern extremes.
}
\label{fig:reg-cuts-rrs}
\end{figure*}

For each of the four subregions of India outlined in Fig.~\ref{fig:corr-maps}, we compute an aggregate cutoff by combining all gridpoints within that region into a single daily rainfall distribution for each summer (Methods).  While the cutoff for Southeast India is negatively correlated with NINO3.4 and marginally statistically significant (\({r=-0.16}\), \(p=0.08\)), the corresponding correlations for All-Monsoonal India, the Central Monsoon Zone, and the Western Ghats regions are all significantly positive (\({r=+0.34}\), \({{+}0.34}\), \({+}0.20\) and \({p=2\times10^{-4}}\), \(2\times10^{-4}\), 0.03, respectively; Methods).  Composites combining either all summers 1901-2020, the 36 \elnino/ summers, or the 41 \lanina/ summers behave similarly: except for Southeast India, the region-aggregated cutoffs are all significantly larger for the \elnino/ than the \lanina/ composites (Fig.~\ref{fig:reg-cuts-rrs}b; Methods).

These signals in extremes differ qualitatively from those in less severe rain rates, which are suppressed throughout the domain by \elnino/.  Pointwise exceedance counts---the number of days in each summer exceeding a specified local rainfall percentile---are predominantly negatively correlated with NINO3.4 for low through moderate percentiles (Extended Data Fig.~\ref{fig:exceed-maps}a-c), becoming positive for the climatologically wet regions only around the 95th percentile and above.  The same overall behavior also holds using fixed accumulations in mm (Extended Data Fig.~\ref{fig:exceed-maps}d-f).

This can also be seen in ENSO risk ratios for each region \cite{neelin_global_2017,martinez-villalobos_why_2019,neelin_precipitation_2022}: at each daily accumulation, the fraction of rainy days exceeding that value in \elnino/ summers divided by the corresponding fraction for \lanina/ summers (Fig.~\ref{fig:reg-cuts-rrs}b,c; Methods).  Accumulations are inhibited under \elnino/ compared to \lanina/ at all percentiles in Southeast India, up to the 75th percentile in the Central Monsoon Zone, and up to the 90th percentile for both the whole domain and the Western Ghats coastal band.  For the latter two especially, the percentile above which likelihoods are enhanced under \elnino/ occurs near that of the climatological cutoff (overlaid circles in Fig.~\ref{fig:reg-cuts-rrs}b), above which probabilities increase quasi-exponentially with rain rate  (Fig.~\ref{fig:reg-cuts-rrs}c)---both signatures of an increased cutoff \cite{neelin_global_2017}.  At the 99.99th rain-intensity percentile, for example, the increases are approximately 43\%, 58\%, and 70\% respectively for the whole domain, Central Monsoon Zone, and Western Ghats.

Cutoffs and risk ratios estimated from gamma fits to each composite distribution yield similar results (Methods; Extended Data Fig.~\ref{fig:gamma-risk-ratios}), as do other metrics of extreme rainfall  (Methods; Extended Data Fig.~\ref{fig:sens-tests}).  These include correlations of block maxima with NINO3.4, pointwise Gumbel fits to the block maxima for \elnino/ \vs/ \lanina/ composites, and quantile regressions of daily rainfall \vs/ JJAS NINO3.4 at the 99th and 99.9th percentiles.  Results are also similar in the Indian Meteorological Department \(1^\circ\times1^\circ\) gridded daily rainfall dataset  \cite{rajeevan_development_2005}, which importantly for extremes \cite{lin_if_2019} is generated from a fixed network of rain gauges (Methods; dashed curves and unfilled elements in Fig.~\ref{fig:corr-maps}, Fig.~\ref{fig:reg-cuts-rrs} and Extended Data Figs.~\ref{fig:mean-vs-n34corr} and \ref{fig:gamma-risk-ratios}).

Finally, while \elnino/'s propensity to generate summer-mean Indian drought notoriously weakened from the pre-satellite (1901-1978) to satellite (1979-2020) eras \cite{kumar_weakening_1999,gershunov_low-frequency_2001,kumar_unraveling_2006}---with respective NINO3.4 correlations \({r=-0.59}\) to -0.41---the ENSO signals in extremes are more robust (Extended Data Fig.~\ref{fig:presat-sat}; Methods).  This holds in terms of the overall geographic patterns seen in pointwise cutoffs, linear regressions of the region-aggregated cutoffs on NINO3.4, and an index linearly combining the region-aggregated cutoffs of the Core Monsoon Zone, Western Ghats, and Southeast India regions (Methods).  Nevertheless, within peninsular India both the increase in extremes with NINO3.4 in the southwestern coastal band and the decrease in the southeast weaken nontrivially.  As such for the region-aggregate cutoffs only the Central Monsoon Zone's regression against NINO3.4 is statistically significant within the satellite era.  The extent to which these modest differences between epochs are physically forced \vs/ merely sampling-driven \cite{gershunov_low-frequency_2001} remains to be seen.

\section*{Mechanisms: ENSO-forced changes in convective buoyancy extremes}

We expect extreme rain where near-surface air is especially warm and moist relative to air above it that, too, is very moist.  These conditions enable the near-surface air to precipitate out copious water as it ascends deeply, despite losing buoyancy to cooling through expansion and to dilution through mixing with the drier ambient air.  These processes have been encapsulated into a scalar metric of convective buoyancy known as \(B_L\) \cite{ahmed_reverse_2018,ahmed_deep_2020}, which we compute for each gridpoint and summer day of 1979 to 2020 using ERA5 reanalysis data \cite{hersbach_era5_2020} (Methods).

\begin{figure*}
\centering
\includegraphics[width=0.5\textwidth]{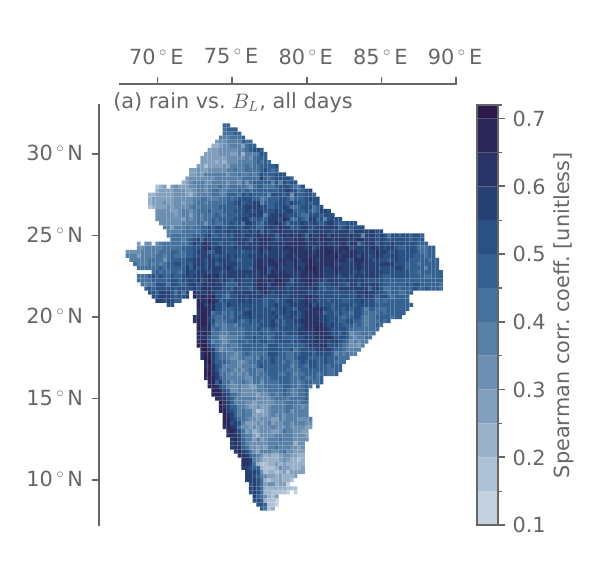}
\includegraphics[width=0.5\textwidth]{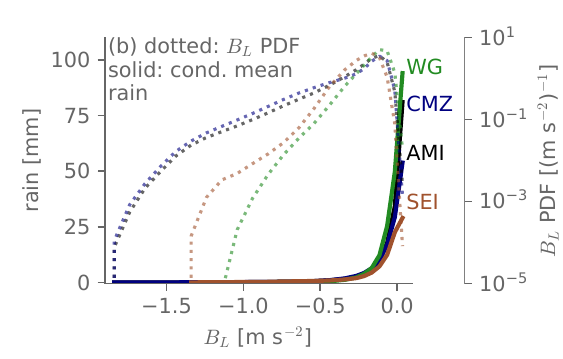}\\
\includegraphics[width=0.5\textwidth]{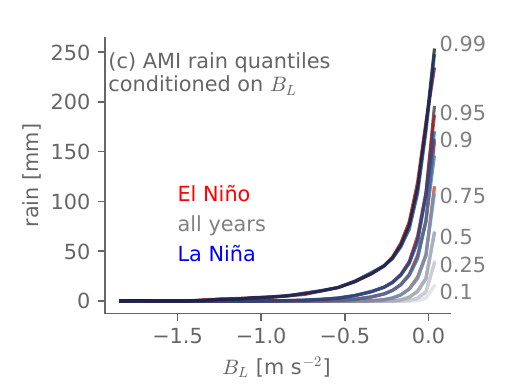}\\
\caption{(a) Spearman correlation coefficient across all JJAS days 1979-2020 at each gridpoint between daily rainfall and the convective buoyancy metric.
  (b) In dotted curves, PDFs of the buoyancy metric for each region.  In solid curves, daily rainfall conditionally averaged on the buoyancy metric. 
  (c) For the whole monsoonal India domain, composite quantiles of rainfall conditioned on the buoyancy metric as labeled.
}
\label{fig:bl-clim}
\end{figure*}

Daily rainfall depends strongly on local convective buoyancy, with nonlinear Spearman correlation coefficient positive at every gridpoint, up to +0.71 (Fig.~\ref{fig:bl-clim}a) and the expected quasi-threshold relationship \cite{ahmed_reverse_2018,ahmed_deep_2020} in region-aggregated averages (Fig.~\ref{fig:bl-clim}b) and quantiles (Fig.~\ref{fig:bl-clim}c) of gridpoint-wise rainfall conditioned on local buoyancy (Methods).  Rainfall is typically absent or weak until buoyancy reaches a threshold value slightly below zero, above which rainfall increases sharply, rapidly consuming buoyancy with it (Fig.~\ref{fig:bl-clim}b).  These conditional curves rise to higher rainfall values in the rainy southwestern coastal band and Central Monsoon Zone compared to the drier southeast India, but to first approximation they are insensitive to the state of ENSO (Fig.~\ref{fig:bl-clim}c).

\begin{figure*}
\centering
\includegraphics[width=\textwidth]{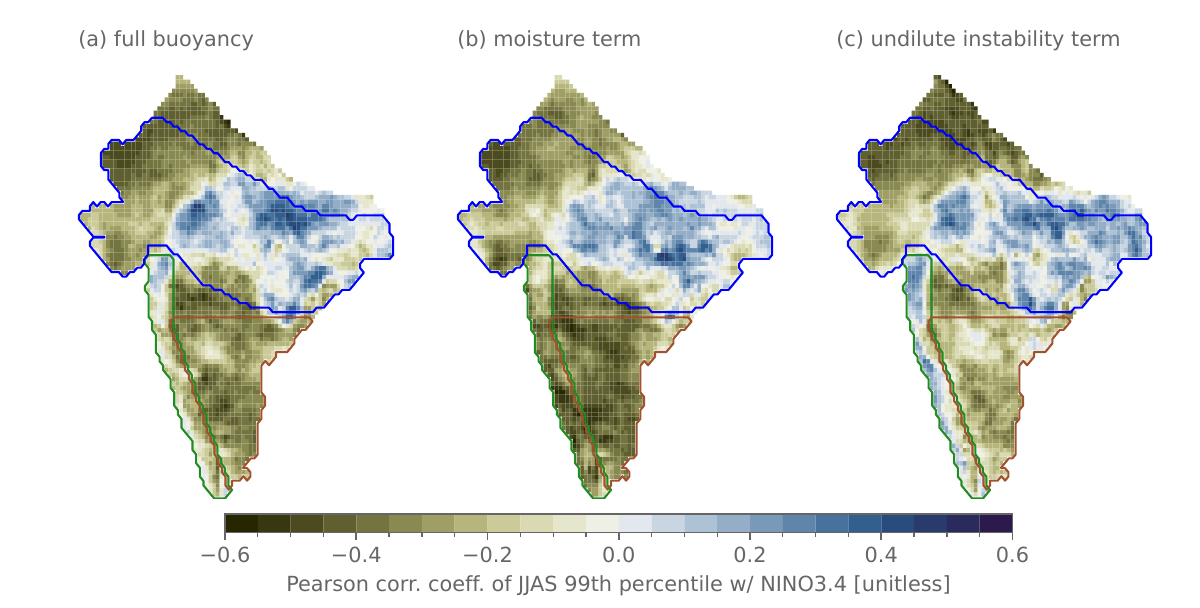}\\
\includegraphics[width=\textwidth]{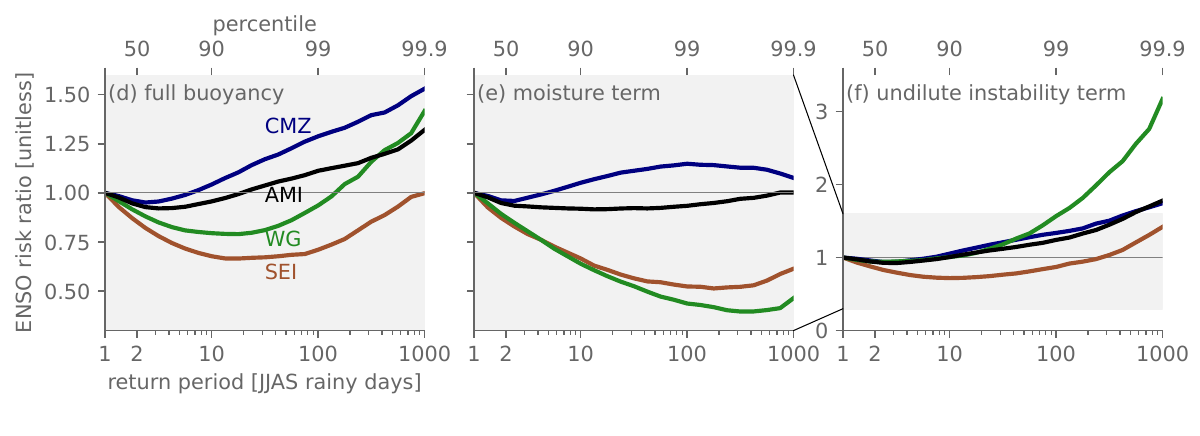}
\caption{
  (a,b,c) Pearson correlation coefficient of NINO3.4 vs. 99th percentile daily values of the (a) full buoyancy metric (b) just the moisture term, or (c) just the undilute instability term for each JJAS.
  (d,e,f) Regional ENSO risk ratios of rainy-day buoyancy, for (d) the full metric, (e) just the moisture term, or (f) just the undilute instability term as a function of return period.
}
\label{fig:bl-enso}
\end{figure*}

Based on correlations at each gridpoint against NINO3.4, summer-mean and moderate-percentile convective buoyancy values are, like rainfall, widely suppressed in \elnino/ summers (not shown).  And also like rainfall, the right-tail values are suppressed in the southeast and northwest but enhanced over much of central-eastern India  (Fig.~\ref{fig:bl-enso}a for 99th percentile values, with results similar for block maxima and other high percentiles, not shown).  The individual buoyancy components of moisture and undilute instability behave similarly to one another over these parts of the domain  (Fig.~\ref{fig:bl-enso}b,c): both very moist and very gravitationally unstable days are predominantly enhanced over eastern-central India \vs/ suppressed over the northwest and southeast.  

What, then, accounts for these convective buoyancy signals?  Preliminary results point toward synoptic low-pressure systems \cite{mooley_aspects_1973}, which generate heavy rainfall within their cores \cite{nikumbh_recent_2019} and track predominantly over Central India, though occasionally traveling farther south or extending to Rajasthan in the northwest \cite{hunt_relationship_2019}.  A prior study indicates that the core of monsoon low-pressure systems are very humid and that during \elnino/ summers the rainfall within those cores is enhanced \cite{hunt_structure_2016}.  And preliminary findings (Yujia You, pers.\@ comm.) indicate that those low-pressure system tracks over and ending in central India are favored by \elnino/, at the expense of those over peninsular India and those long tracks reaching northwestern India.  If true, this would promote very humid, buoyant days in central-eastern India and suppress them in the southeast and northwest.

For the southwestern coastal band, which is less directly influenced by these low pressure systems, nearly saturated days above the boundary layer are strongly suppressed in \elnino/ summers while highly unstable days are strongly enhanced.  This undilute buoyancy signal emerges to some degree for the other regions as well, as is particularly evident in region-aggregated ENSO risk ratios of the convective buoyancy terms on rainy days (Fig.~\ref{fig:bl-enso}d-f; Methods).  Moreover, based on differences in the ENSO composites of rainfall conditioned on the buoyancy terms (not shown), tropospheric dryness inhibits heavy rainfall somewhat less during \elnino/ summers; this holds for each region but especially the southwest.  We therefore speculate that the anomalous free-tropospheric subsidence and dryness under \elnino/ throughout the domain (Extended Data Fig.~\ref{fig:enso-mean}) suppress moderate rain events, enabling higher undilute instability values to occasionally build up, fueling extreme convection once a passing wave or advective event re-moistens the lower free troposphere just enough.  This mechanism would be broadly consistent with observational and modeling evidence for less frequent but more intense rain events in drier ambient conditions \cite{raymond_rains_2024}.  Moreover, though it would in principle operate throughout the domain, plausibly it would be strongest in the southwest, where the prevailing monsoon southwesterlies reliably converging warm air into the boundary layer, promoting buoyancy.

\section*{Outlook: seasonal forecasts, climate change, and other regions}

Despite ENSO's known ``spring predictability barrier'' \cite{tippett_excessive_2020}, existing seasonal forecasts of ENSO during boreal summer are skillful \cite{tippett_assessing_2019}; for example, the multi-model-mean forecasted JJAS value of NINO3.4 from the North American Multi-Model Ensemble (NMME) \cite{kirtman_north_2014} seasonal forecast integrations initialized each May 1st are correlated with the actual value at \(r=0.86\) over 1981-2020 (Methods).  We have re-computed the correlations between the JJAS cutoff rain rate and NINO3.4 using this forecasted value, and the overall signals remain (Extended Data Fig.~\ref{fig:nmme}).  This preliminary result raises the prospect of skillful ENSO-based seasonal forecasts of extreme rainfall probability within monsoonal India. 

An observed, long-term trend toward more extreme daily rainfall events in central India into the 2000s has been argued to stem from mean moistening of the troposphere due to global warming \cite{goswami_increasing_2006,rajeevan_analysis_2008}, as have future projections of increased extreme rainfall events globally \cite{neelin_precipitation_2022}.  Notwithstanding that more recent trends in the Indian summer monsoon have been flat or negative \cite{bajrang_possible_2023}, underlying this argument are the assumptions that right-tail daily moisture events (a) track with the season-mean value and (b) predominate over changes in undilute instability.  Based on our results, however, for ENSO (a) is violated within central India and (b) is violated within the southwestern coastal band---precisely the regions where daily extreme rainfall is projected to increase most in model simulations under future global warming \cite{katzenberger_robust_2021}.  This argues for revisiting both observed trends and model projections of extreme rainfall, incorporating the full daily distributions of rainfall and of the convective buoyancy terms.  Only then can we meaningfully evaluate whether these observed, interannual, internally generated ENSO signals constitute a meaningful emergent constraint on model-projected, secular, externally forced trends \cite{klein_emergent_2015}.

Our approach could also be applied to climate variability modes in the Indian \cite{ajayamohan_influence_2008,ajayamohan_indian_2008} and Atlantic Oceans \cite{borah_indian_2020} thought to influence the summer monsoon, as well as to hourly rainfall.  Many Indian summer monsoon rain events span considerably less than 24 hours \cite{moron_storm_2021}, and a study \cite{li_global_2020} using hourly rainfall suggests that the average rain rate during rainy hours increases with \elnino/ over much of the Tropics.  Perhaps, as for daily rain in the Indian summer monsoon, these signals reflect buoyancy-driven increased extremes.

\bibliography{./references.bib}

\newpage
\section*{Methods}

\subsection*{Region definitions}
The four subregions of India are those used by \cite{hill_all-india_2022}.  The ``All Monsoonal India'' (AMI) region comprises all gridpoints in the monsoonal India domain defined by \cite{gadgil_new_2019}.  The remaining three are non-overlapping subsets of AMI.  The Central Monsoon Zone (CMZ) definition also follows \cite{gadgil_new_2019}, spanning from Pakistan to the Bay of Bengal in a wide band north of peninsular India.  The Western Ghats (WG) region is defined as all land points within the polygon whose vertices are (76.5\(^\circ\)E, 7.5\(^\circ\)N), (78.5\(^\circ\)E, 7.5\(^\circ\)N), (74\(^\circ\)E, 18\(^\circ\)N), (74\(^\circ\)E, 21.5\(^\circ\)N), (72.25\(^\circ\)E, 21.5\(^\circ\)N), and (72.25\(^\circ\)E, 18\(^\circ\)N).  The Southeast India (SEI) region includes all points south of 18\(^\circ\)N that are not within WG.  The WG and SEI region borders were defined by \cite{hill_all-india_2022} explicitly based on JJAS-mean rainfall properties in the Indian Meteorological Department 0.25\(^\circ\) dataset to distinguish the coherent regimes within peninsular India, but neither they nor the other regions were defined explicitly with reference to extreme rainfall characteristics.

Region-aggregated distributions are computed by combining fields from all gridpoints of a given region into a single distribution without any area weighting.  Region-averaged fields are area-weighted averages across all gridpoints within each region.  

\subsection*{Indian Meteorological Department \({0.25^\circ\times0.25^\circ}\) rainfall dataset}
Daily rainfall data comes from the 0.25\(^\circ\times\)0.25\(^\circ\) latitude-longitude gridded product produced by the Indian Meteorological Department (IMD) \citep{pai_development_2014}.  This dataset spans all days of the year from January 1, 1901 to the present, and we use all days in June, July, August, and September (JJAS) from 1901 through 2020.  This gridded product is derived from a dense network of in situ rain gauges with coverage throughout India, using a simple interpolation procedure to go from the irregular station coverage to the regular latitude-longitude grid.  The rain gauge coverage varies in time, with \(\sim\)1500 stations in 1901, peaking above 4000 stations in the early 1990s, and then decreasing appreciably to \(\sim\)2000 in the 2010s \citep{pai_development_2014}.  These secular trends in station coverage have been argued to strongly influence inferences about long-term trends in daily rainfall extremes \citep{lin_if_2019}.  For our purposes, however, for interannual variability it is unlikely that the fluctuations in station coverage would alias onto ENSO.

This dataset contains spurious values in a few gridpoints and times, which appear to us to stem from errant reported values from one or more stations which then propagate in space through the interpolation procedure.  In particular, in northeast India directly along the border with Bangladesh, starting in the 1970s values become insensible.  We mask these out along with the rest of northeastern and far northern India as part of our ``monsoonal India'' mask \citep{gadgil_new_2019}.  Despite these problems, the density of rain gauges dating back to the 19th century and the stewardship of the resulting data by the Indian Meteorological Department make this a very high-quality dataset.

\subsection*{Cutoff rain rate}
We compute the cutoff daily rain rate using the method of moments, which for a given daily rainfall distribution is simply the variance divided by the mean.  This measure is motivated by the correspondence of daily rainfall distributions to gamma distributions \cite{martinez-villalobos_why_2019}.  In terms of shape parameter \(k\) and scale parameter \(\theta\), the gamma probability density function, \(f(x)\), follows \(f(x)\propto x^{k-1}e^{-x/\theta}\), from which the cutoff is easily identified as the scale parameter \(\theta\).  Moreover, for gamma distributions the variance is exactly \(k\theta^2\) and the mean is \(k\theta\).  Thus, insofar as the daily rainfall distributions approximately follow gamma distributions, the variance-to-mean ratio usefully estimates the cutoff.

The method of moments is easily calculated, but being computed over the entire distribution, it is influenced by behavior at the low end of the daily rainfall distribution as well as the extremes.  Given the strong influence of ENSO on no- and low-rainfall day frequencies, this potentially makes it a biased metric of extremes.  We have tested the sensitivity to behaviors at low rain rates by computing it restricting to rain rates over different thresholds.  This increases the mean of the remaining sample, yielding a low-biased cutoff estimate, but results in terms of correlations with NINO3.4 or ratios of ENSO composite quantities are qualitatively insensitive (not shown).

An alternative method for computing the cutoff is via a linear regression of the daily rainfall PDF in semi-log space \cite{dulguerov_extreme_2022}.  Denoting the daily rainfall PDF as \(f(P)\), a linear regression is performed of \(\log(f(P))\) on \(P\) but restricting to the right tail of the distribution (which, being quasi-exponential, is close to a straight line); denoting the slope of this regression \(r\), the cutoff estimate is then \(-1/r\).  While this avoids contamination by the low end of the distribution, we find it strongly determined by the few largest values: for the parameter choices explored, we find the regional cutoffs for each JJAS computed via the regression to all be correlated at \(r>0.9\) with the region's JJAS block maximum rain, compared to \(r\sim0.5-0.7\) with the method of moments.

As such, we elect to use the cutoff computed from the method of moments, taking it to be more representative of behaviors throughout the right tails of the daily rainfall distributions.

\subsection*{NINO3.4 index}
As our index of ENSO, we compute the standard NINO3.4 index of sea surface temperature (SST) anomalies averaged over the central and eastern equatorial Pacific: 120\(^\circ\)–170\(^\circ\)W, 5\(^\circ\)S–5\(^\circ\)N.  The monthly SSTs come from the National Oceanic and Atmospheric Administration Extended Reconstructed SST (ERSST) dataset, version 5 \citep{huang_extended_2015}.  The SSTs are first averaged with area-weighting over the NINO3.4 region for each month.  They are then averaged over JJAS for each year.  The linear trend computed by least squares regression over 1901-2020 of this timeseries of JJAS values is then computed and subtracted off.  Finally, the time mean of this detrended timeseries is computed and subtracted off to generate an anomaly timeseries; the result is the NINO3.4 index we use.

We have also repeated key calculations using JJAS deviations of the NINO3.4 averaged SSTs from a 30-year rolling mean with no detrending, more akin (but not identical) to the NOAA Oceanic Ni\~no Index (ONI), finding little sensitivity.  For example, the region cutoffs for each JJAS are correlated with ONI at \({r=0.29}\), 0.29, 0.19, and -0.19 for AMI, CMZ, WG, and SEI, respectively, all less positive but within 0.05 of the corresponding correlations using NINO3.4.

\subsection*{Detrending}
The trend for each field is computed via least-squares regression in time over 1901-2020.  De-trended fields are constructed by subtracting this trend from the full field.  All correlation coefficients are computed on fields linearly detrended in this way.

\subsection*{Statistical significance estimates}
We use a standard bootstrapping technique to assess the statistical significance of the pointwise correlations between the cutoff and NINO3.4.  Specifically, we randomly draw with replacement 120 years from the 1901-2020 period, taking the NINO3.4 value and each point's cutoff for that JJAS.  We then compute the Pearson correlation coefficient at each gridpoint for that randomly drawn sample of 120 years.  We repeat this procedure 10,000 times, and finally construct a 95\% confidence interval by selecting the 2.5th and 97.5th percentiles for each gridpoint of these 10,000 correlation coefficients.  The inferred confidence intervals converge rapidly as the number of bootstrap draws exceeds 100 (not shown).  Those points whose confidence interval do not cross zero we deem statistically significant at the 95\% confidence level.  

For the ENSO-composite region-aggregated cutoffs, we proceed as follows.  First, we compute region-aggregated cutoff rain rates for each individual JJAS.  For both ENSO composites and for the composite over all years 1901-2020, the average across these yearly cutoffs is similar to the cutoffs computed across all years (compare filled circles to center of corresponding vertical lines in Fig.~\ref{fig:reg-cuts-rrs}b).  We therefore use the sample sizes (36 \elnino/ years, 41 \lanina/ years, 120 total years) and the interannual standard deviations in the yearly cutoffs for each region and ENSO composite to construct the 95\% confidence intervals (vertical lines in Fig.~\ref{fig:corr-maps}b).  These regional yearly cutoffs are reasonably normally distributed (not shown), and so we employ a standard $t$-test to assess the statistical significance of the correlations between the yearly cutoffs and NINO3.4.  The resulting \(p\) values are \(p=0.0002\), 0.0003, 0.03, and 0.06 for AMI, CMZ, WG, and SEI, respectively.

\subsection*{ENSO composites and risk ratios}
For ENSO-based composites, we use a \(\pm0.3\)~K threshold of the NINO3.4 index: all summers w/ JJAS NINO3.4 exceeding +0.3~K are designated \elnino/, those between -0.3 and +0.3 neutral, and those less than -0.3 \lanina/.  These -0.3 and +0.3 threshold values correspond to the 34th and 70th percentiles of the JJAS NINO3.4 distribution respectively.  Results appear insensitive to this threshold from values of 0 to roughly \(\pm1.0\).  For example, the AMI composite cutoffs for a 0~K threshold are 36.7 and 33.8~mm respectively for \elnino/ and \lanina/; for a \(\pm1\)~K threshold they are 36.4 and 33.5~mm, respectively, with only small fluctuations for intermediate values.  The \(\pm0.3\)~K threshold was subjectively chosen as a compromise between maximizing the sample size of each composite \vs/ maximizing the difference in the large-scale atmospheric state between the composites.

To compute the ENSO risk ratio for a given field, first we compute the empirical cumulative distribution function (CDF, denoted \(F(x)\)) for that field for the \elnino/ composite and separately for the \lanina/ composite.  By definition of the CDF, in either composite, the empirical probability of exceeding a given value equals one minus the CDF at that value: \(P(X\geq x)=1-F(x)\).  As such, the ENSO risk ratio is computed as \((1-F_\mr{EN}(x))/(1-F_\mr{LN}(x))\), where the subscripts ``EN'' and ``LN'' refer to \elnino/ and \lanina/ respectively.

We compute risk ratios in this manner for daily rain accumulation on rainy days (using a standard \({>}1\)~mm threshold).  We then interpolate these risk ratios, which are functions of daily rain accumulation in mm, to climatological quantiles of daily rainfall across all years 1901-2020 and all points in the given region.  Denoting a given quantile \(q\), the quantiles are converted as standard to return periods as \(1/(1-q)\).  Because the risk ratios restrict to rainy days, the return periods correspond to JJAS rainy days as well rather than JJAS calendar days.

\subsection*{Exceedance counts}
The exceedance count is defined for each gridpoint, year, and daily rainfall percentile as the number of days in that year's JJAS and gridpoint exceeding the climatological rainfall at the given percentile across all years and all days of JJAS.  For example, if a given year and gridpoint experienced rainfall exceeding the climatological 90th percentile on three days, the exceedance count at the 90th percentile would be three.

The exceedance count can also be applied to fixed \mmday/ thresholds rather than local percentiles.  Extended Data Fig.~\ref{fig:exceed-maps}d,e,f shows the correlations \vs/ NINO3.4 of the yearly exceedance counts for three of the official thresholds in \mmday/ the Indian Meteorological Department uses to categorize rainfall severity.

\subsection*{Explicit gamma fits}
For each region and each ENSO-composite daily rainfall distribution, a gamma distribution was fitted for all daily accumulations \(>0.1\)~mm using the \texttt{scipy.stats.gamma.fit} function from the SciPy package \cite{virtanen_scipy_2020} for Python.  This yields for each distribution values of the three parameters of the gamma distribution, location, shape, and scale.  In the exact gamma limit, the cutoff as estimated by the method of moments is identical to the scale parameter, and we therefore take the scale parameter for each of these fits as the gamma-estimated cutoff.  The ENSO risk ratios computed for the explicit gamma fits were computed from their CDFs identically to those for the empirical distributions.

\subsection*{Other extreme daily rainfall metrics}
The block maximum daily rainfall for each JJAS is simply the single daily maximum rainfall in that JJAS at each gridpoint.  We create \elnino/ and \lanina/ composite block-maxima distributions for each gridpoint by combining the block maxima across all \elnino/ summers or all \lanina/ summers into a single distribution.  We then fit a Gumbel distribution to each \elnino/ composite and \lanina/ composite using the \texttt{scipy.stats.gumbel\_r.fit} function, yielding for each gridpoint and composite a scale parameter value and a location parameter value.  Increasing either the scale or location parameter of the Gumbel distribution increases right-tail probabilities.

Quantile regression determines the linear slope of the dependent variable (in this case, daily rainfall) \vs/ the independent variable (in this case, the JJAS NINO3.4 value) such that, for the specified quantile \(q\), a fraction \(q\) of the dependent variable points fall below that line and a fraction \(1-q\) of points fall above that line \cite{mckinnon_changing_2016}.  For a given JJAS, each day is assigned the JJAS-mean NINO3.4 value, and for each gridpoint all JJAS days of all years are combined into one distribution.  We perform quantile regression for the 99th and 99.9th percentiles.

\subsection*{Indian Meteorological Department \({1^\circ\times1^\circ}\) rainfall dataset}
We have replicated all key analyses on the Indian Meteorological Department's 1\(^\circ\times\)1\(^\circ\) gridded product \citep{rajeevan_development_2005}.  This product uses the same interpolation procedure of Indian Meteorological Department station data, but in addition to being coarser, it differs from the 0.25\(^\circ\) product in that its network of 1,803 stations is fixed in time.  There are no substantive discrepancies in our key findings between the two products.  Perhaps most notable is that daily rainfall values within the very narrow WG coastal band reach considerably larger values in the higher-resolution product, which is reflected in the WG region-aggregated daily cutoff (Fig.~\ref{fig:reg-cuts-rrs}b).  But this amounts to a approximately constant offset; the ENSO signals in this and other regions are similar between the two products.

\subsection*{Pre-satellite and satellite era results}
A multiple linear regression over 1901-2020 on NINO3.4 of the Central Monsoon Zone, Western Ghats, and Southeast India cutoffs yields respective weights of +1.0, +0.4, and -1.0, and the resulting timeseries is correlated with NINO3.4 for the full, pre-satellite, and satellite periods respectively at \({r=0.45}\), 0.46, and 0.44.  In the satellite era, the JJAS NINO3.4 \(\pm\)0.3~K threshold results in 14 \elnino/ summers and 13 \lanina/ summers.

\subsection*{ERA5 reanalysis data}
Large-scale environmental fields come from the European Center for Medium-range Weather Forecasting (ECMWF) ERA5 reanalysis dataset \citep{hersbach_era5_2020}, which provides data at hourly and monthly resolutions on a global 0.25\(\times\)0.25\(^\circ\) grid.  Conveniently, over India this grid aligns identically with that of the Indian Meteorological Department rainfall.  Though the ERA5 dataset now extends from 1950 to present, we restrict to the satellite era, 1979-2020, as the amount of observational data assimilated into the underlying numerical model increases dramatically with the availability of satellite retrievals.  We compute JJAS-mean quantities from ERA5 monthly fields for June through September, and we compute daily-mean fields from ERA5 hourly fields, averaging over all 24 hourly fields.

\subsection*{\(B_L\) convective buoyancy metric}
We compute the convective buoyancy metric, \(B_L\), following \cite{ahmed_deep_2020} as
\begin{equation}
  \label{eq:bl-def}
  B_L\equiv g\left[\wpbl\frac{\thetaepbl-\thetaesatlft}{\thetaesatlft}+\wlft\frac{\thetaelft-\thetaesatlft}{\thetaesatlft}\right],
\end{equation}
where \(g\) is gravity, \(\theta_e\) is equivalent potential temperature, \(\thetaepbl\) is \(\theta_e\) averaged over the boundary layer, \(\theta_e^*\) is saturation equivalent potential temperature, \(\thetaelft\) is \(\theta_e\) averaged over the lower free troposphere, \(\thetaesatlft\) is \(\theta^*_e\) averaged over the lower free troposphere, and \(\wpbl\) and \(\wlft\) are the weights given to the undilute instability and subsaturation terms respectively.  Equivalent potential temperature is computed conventionally as
\begin{equation}
  \label{eq:theta-e}
  \theta_e\equiv T\left(\frac{p_0}{p_\mr{d}}\right)^{R_\mr{d}/c_p}\mathcal{H}^{-R_\mr{v}r_\mr{v}/c_p}\exp\left(\frac{L_\mr{v}r_\mr{v}}{c_p T}\right),
\end{equation}
where \(T\) is temperature, \({p_0=1000}\)~hPa is a reference pressure, \(p_\mr{d}\) is the dry air partial pressure, \(R_\mr{d}\) is the dry air gas constant, \(c_p\) is the specific heat of dry air at constant pressure, \(\mathcal{H}\) is relative humidity, \(R_\mr{v}\) is the gas constant of water vapor, and \(r_\mr{v}\) is the water vapor mixing ratio.  This definition neglects the contribution to \(\theta_e\) of liquid water.  Saturation equivalent potential temperature is computed using \eqref{eq:theta-e} but setting \({\mathcal{H}=1}\) and the vapor mixing ratio to its saturation value.  

The layer weights are given by
\begin{align}
  \label{eq:wb-def}
  \wpbl&\equiv \frac{a\Delta p_\pbl}{b\Delta p_\lft}\ln\left(\frac{a\Delta p_\pbl + b\Delta p_\lft}{a\Delta p_\pbl}\right)\\\
  \wlft&\equiv 1 - \wpbl,
\end{align}
where \(a\) and \(b\) are parameters relating to the relative mass inflow rates in the PBL and LFT layers, respectively, and \(\Delta p_\pbl\) and \(\Delta p_\lft\) are the pressure thicknesses of the PBL and LFT layers, respectively.  We follow \cite{ahmed_deep_2020} in setting \(a=b=0.5\), but we compute \(\Delta p_\pbl\) and \(\Delta p_\lft\) for each day and gridpoint as follows.

Following convention \cite{ahmed_deep_2020}, we define the boundary layer to extend from the local surface pressure to 150 hPa above. 
To better resolve the layer boundaries and thus depths, ERA5 fields are linearly interpolated in pressure from their native resolution (of 25~hPa spacing from 1000 to 750 hPa, 50~hPa spacing from 750 to 250 hPa, and 25~hPa spacing from 250 to 100 hPa) to 5~hPa spacing up to 200 hPa.  The LFT is taken to span from the next level vertically above to the last level below the freezing level, defined as where the temperature drops below 273~K.

When computing the free-tropospheric subsaturation term, \(\blsubsat\), and the subcloud undilute instability term, \(\blinstab\), individually, the layer-depth weights are included in each:
\begin{align}
  \label{eq:bl-terms}
  \blsubsat&\equiv g\wlft\frac{\thetaepbl-\thetaesatlft}{\thetaesatlft}\\
  \blinstab&\equiv g\wpbl\frac{\thetaelft-\thetaesatlft}{\thetaesatlft}.
\end{align}

Given our focus on extremes, we use this version of \(B_L\) in terms of equivalent potential temperatures rather than the simplified version in terms of moist enthalpies \cite{ahmed_deep_2020}, as the latter loses some of the variations within either tail.  Our use of a spatiotemporally varying PBL bounds and LFT layer thickness defined in terms of the local surface pressure and freezing level are also more involved than the original version which uses constant values of 1000~hPa as the surface pressure, 850~hPa as the PBL top, and 500~hPa as the LFT top.

\subsection*{ENSO risk ratios of \(B_L\)}
Risk ratios for \(B_L\), \(\blsubsat\), and \(\blinstab\) components are computed identically to those for rainfall, other than being restricted to the satellite era 1979-2020.  This includes restricting to rainy days and the risk ratios then being interpolated to regional climatological quantiles and then return periods.

\subsection*{Data availability statement}
Indian Meteorological Department rainfall data used is available at \url{https://www.imdpune.gov.in/cmpg/Griddata/Rainfall_25_NetCDF.html} and \url{https://www.imdpune.gov.in/cmpg/Griddata/Rainfall_1_NetCDF.html}.  ERSST data is available at \url{https://www.ncei.noaa.gov/pub/data/cmb/ersst/v5/netcdf/}.  ERA5 data is available at \url{https://www.ecmwf.int/en/forecasts/dataset/ecmwf-reanalysis-v5}.  NMME data is available at \url{https://iridl.ldeo.columbia.edu/SOURCES/.Models/.NMME/}.


\backmatter



\newpage
\bmhead{Acknowledgments}

S.A.H. acknowledges funding from NSF award AGS-2123327.  S.A.H., M.B., M.A.C., and A.H.S. acknowledge support from the Monsoon Mission Project under India’s Ministry of Earth Sciences.

\bmhead{Author contributions}
S.A.H. designed the study, oversaw preliminary analyses performed by D.M.Z., performed most analyses, and led writing the manuscript.  D.M.Z. performed the initial analyses establishing the ENSO-extreme rainfall signals.  A.H.S., M.B., and M.A.C. provided scientific feedback throughout the project.   M.K.T. provided guidance on statistical procedures.  F.A. provided guidance on convective buoyancy.  All authors contributed to revising the manuscript.

\bmhead{Competing interests}
The authors declare no competing interests.

\bmhead{Materials \& Correspondence}
All correspondence and material requests should be addressed to S.A.H.

\newpage
\section*{Extended Data Figures}

\renewcommand{\figurename}{Extended Data Fig.}
\setcounter{figure}{0}

\begin{figure*}[htbp]
\centering
\includegraphics[width=0.8\textwidth]{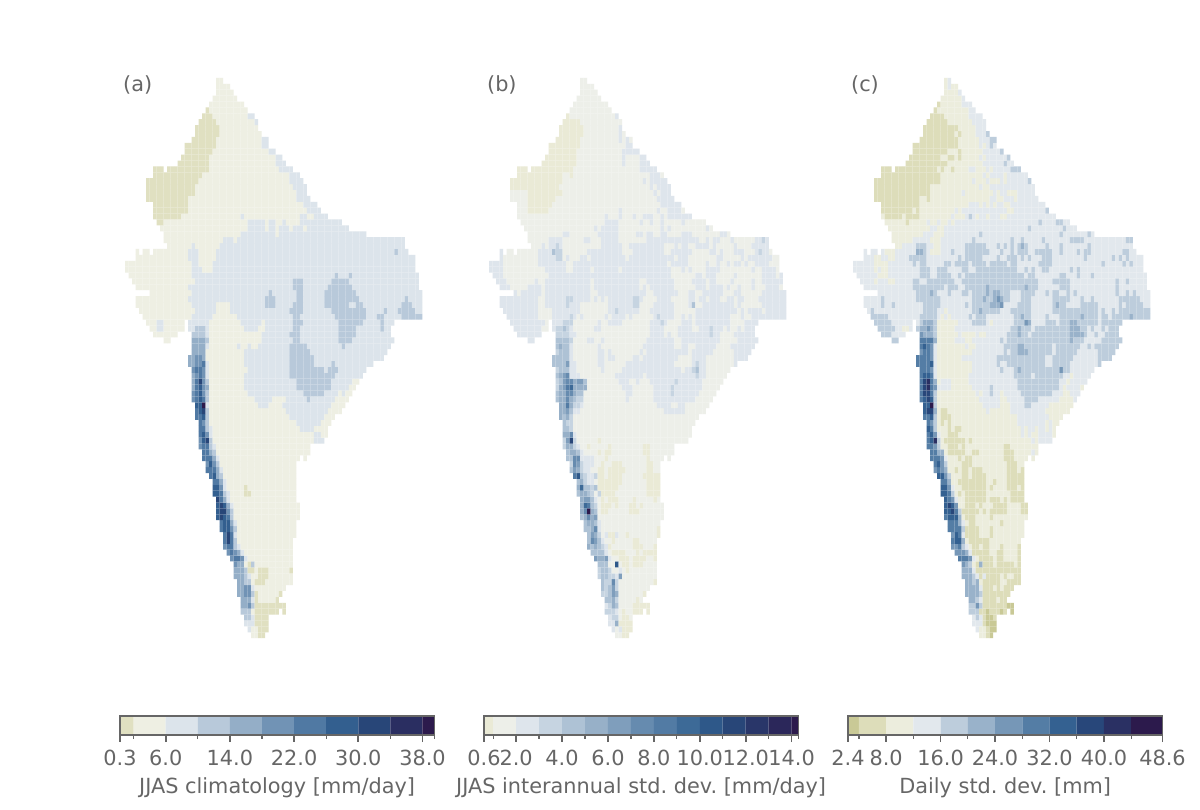}\\
\includegraphics[width=0.8\textwidth]{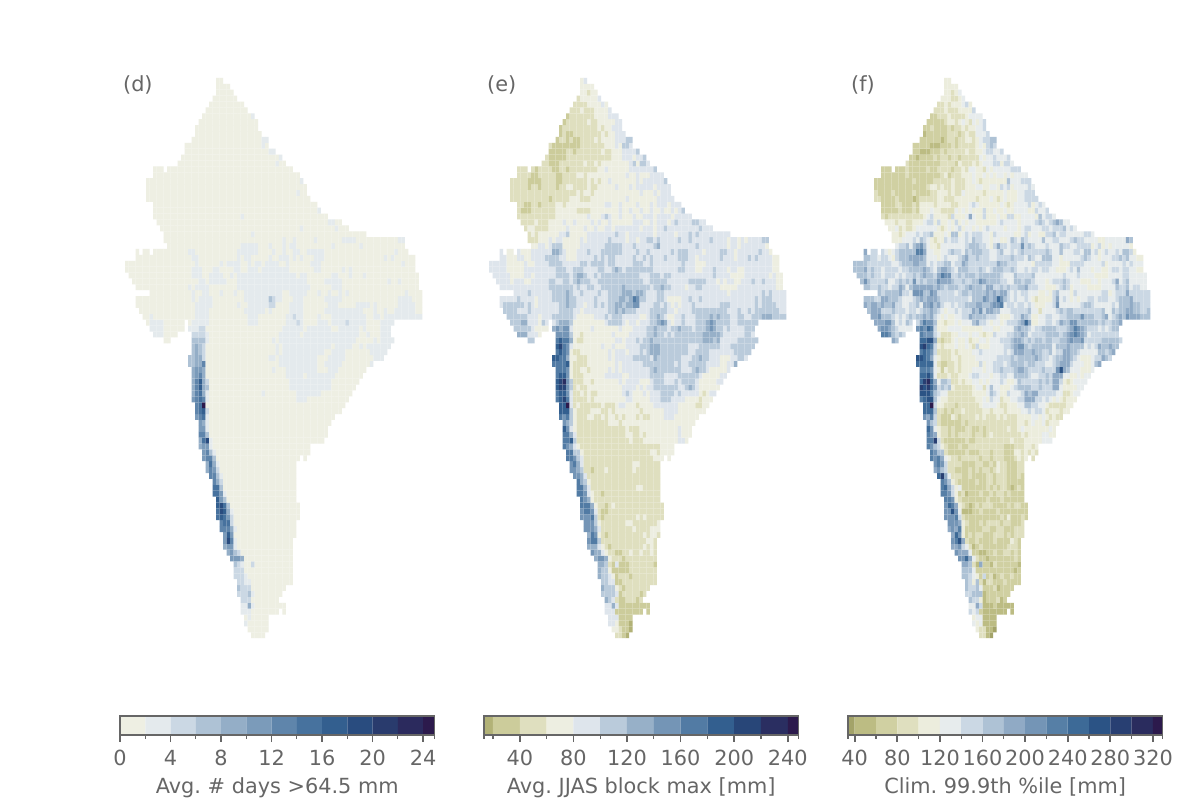}
\caption{In color shading according to each colorbar, (a) climatological JJAS-mean rainfall (units mm/day), (b) JJAS-mean rainfall interannual standard deviation (units mm/day), (c) daily rainfall standard deviation (units mm), (d) average number of days exceeding the Indian Meteorological Department's ``heavy rain'' threshold of 64.5~mm, (e) average across years of each year's JJAS block maximum rain (units mm), and (f) the climatological 99.9th percentile daily accumulation across all JJAS days (units mm).  For each, brown shades denote values less than approximately the median across gridpoints, and blue shades denote values above that value.}
\label{fig:clim-and-stdevs}
\end{figure*}

\newpage
\begin{figure*}[htbp]
\centering
\includegraphics[width=0.5\textwidth]{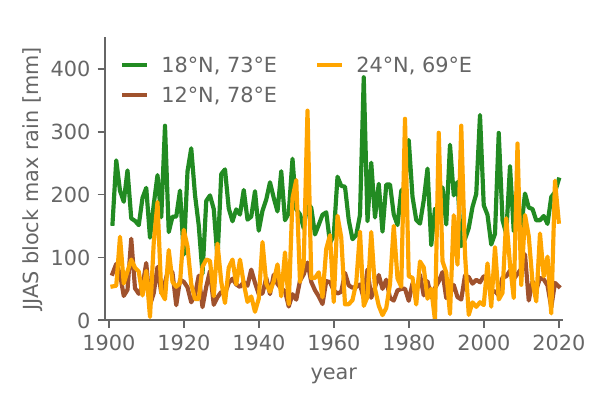}
\includegraphics[width=0.2\textwidth]{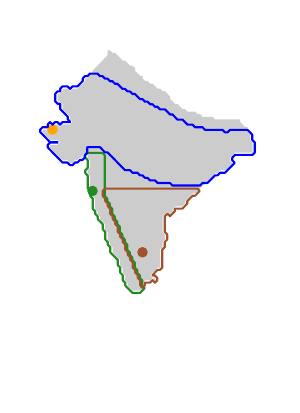}
\caption{Timeseries of daily rainfall block maximum within each JJAS at three representative points.  In very rainy points of WG (dark green), there are very heavy rain events nearly every summer; in dry points of SEI (brown), the heaviest events are much weaker.  In Gujarat (far western CMZ; orange), most summers see modest max rainfall but occasionally receive extremely high values.}
\label{fig:3points}
\end{figure*}

\newpage
\begin{figure*}
\centering
\includegraphics[width=0.5\textwidth]{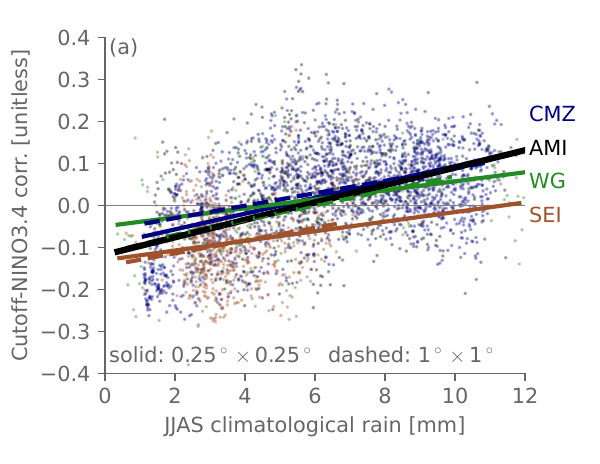}
\caption{Scatterplot of the JJAS cutoff-NINO3.4 correlation coefficient \vs/ climatological JJAS-mean rainfall for points with JJAS climatology \(<12\)~\mmday/ restricting to points below the JJAS climatological rain 95th percentile across gridpoints (12.0~\mmday/).  Overlaid lines are the corresponding linear regressions, from the \(0.25\times 0.25^\circ\) dataset in solid and the \(1^\circ\times 1^\circ\) dataset in dashed.
}
\label{fig:mean-vs-n34corr}
\end{figure*}

\newpage
\begin{figure*}
\centering
\includegraphics[width=\textwidth]{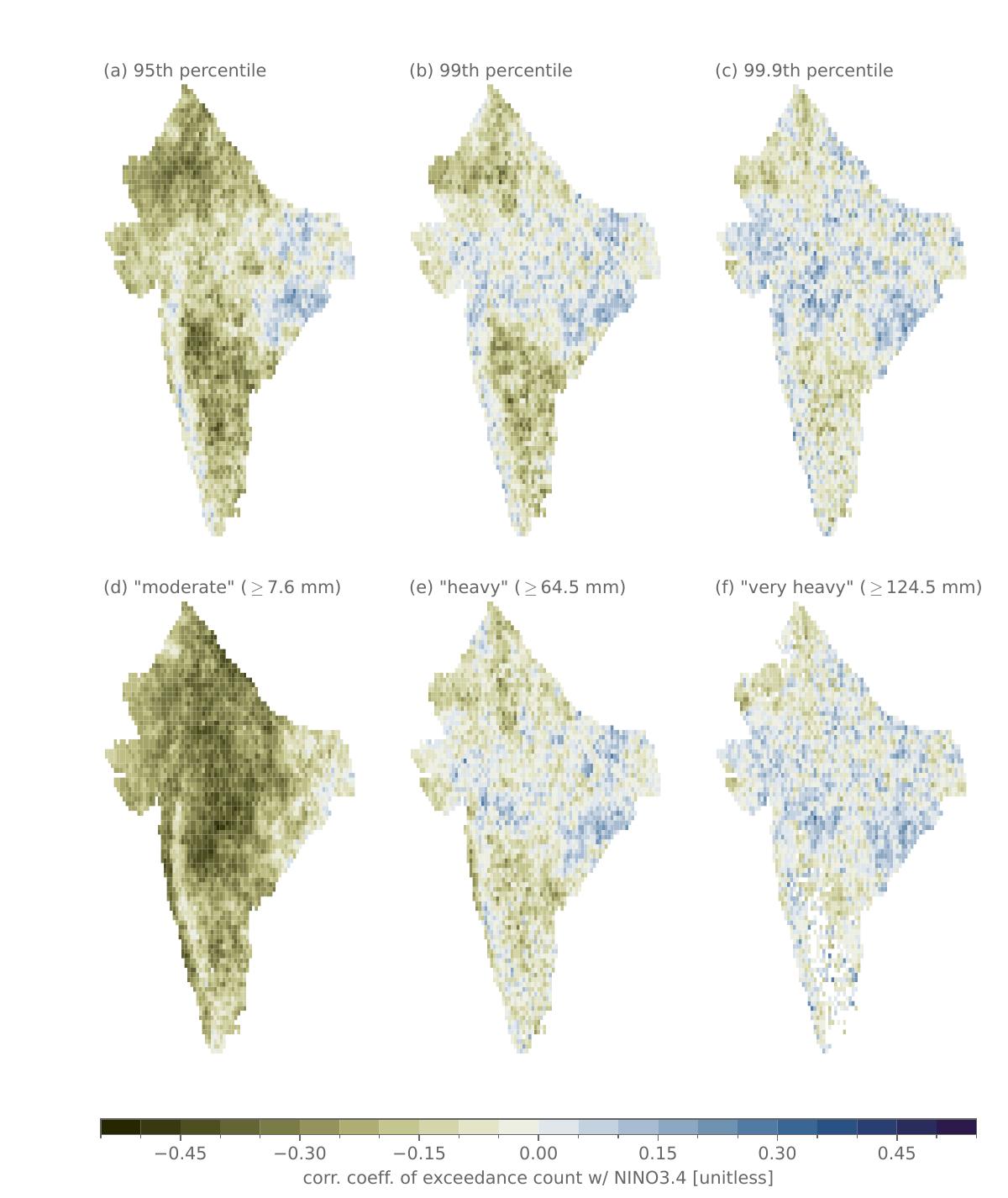}
\caption{In color shading according to the colorbar, correlation coefficient of exceedance counts for each JJAS against NINO3.4 of the local climatological (a) 95th, (b) 99th, and (c) 99.9th percentile, or of the Indian Meteorological Department's thresholds for (d) moderate, (e) heavy, or (f) very heavy daily accumulations, with the corresponding value in mm printed in each label.}
\label{fig:exceed-maps}
\end{figure*}

\begin{figure*}
\centering
\includegraphics[width=0.49\textwidth]{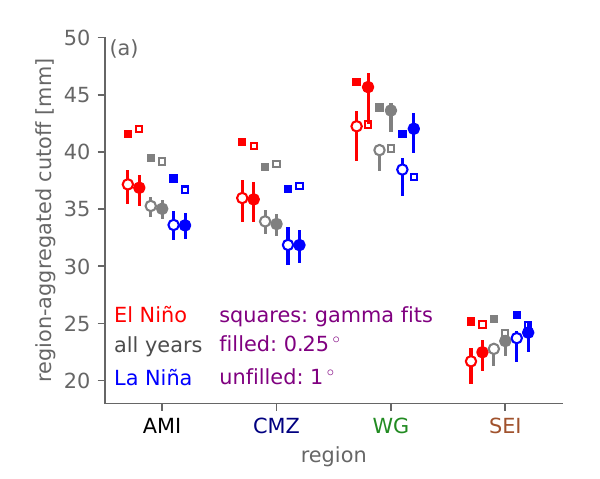}
\includegraphics[width=0.49\textwidth]{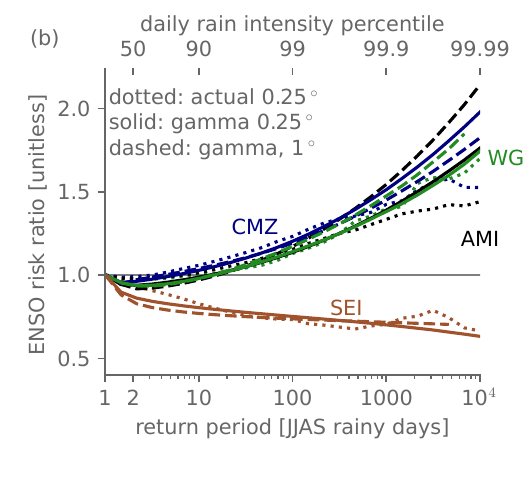}\\
\caption{(a) In squares, composite regional cutoffs computed from gamma fits; all other plotted elements are identical to Fig.~\ref{fig:reg-cuts-rrs}(a).  (b) ENSO risk ratios of daily rainfall for each of the four monsoonal India regions according to the colors and labels.  Dotted curves are the actual risk ratios from the $0.25^\circ\times0.25^\circ$ Indian Meteorological Department rainfall dataset.  Solid curves are the corresponding gamma fits.  Dashed curves are the gamma fits for the $1^\circ\times1^\circ$ dataset.}
\label{fig:gamma-risk-ratios}
\end{figure*}

\newpage
\begin{figure*}
\centering
\includegraphics[width=\textwidth]{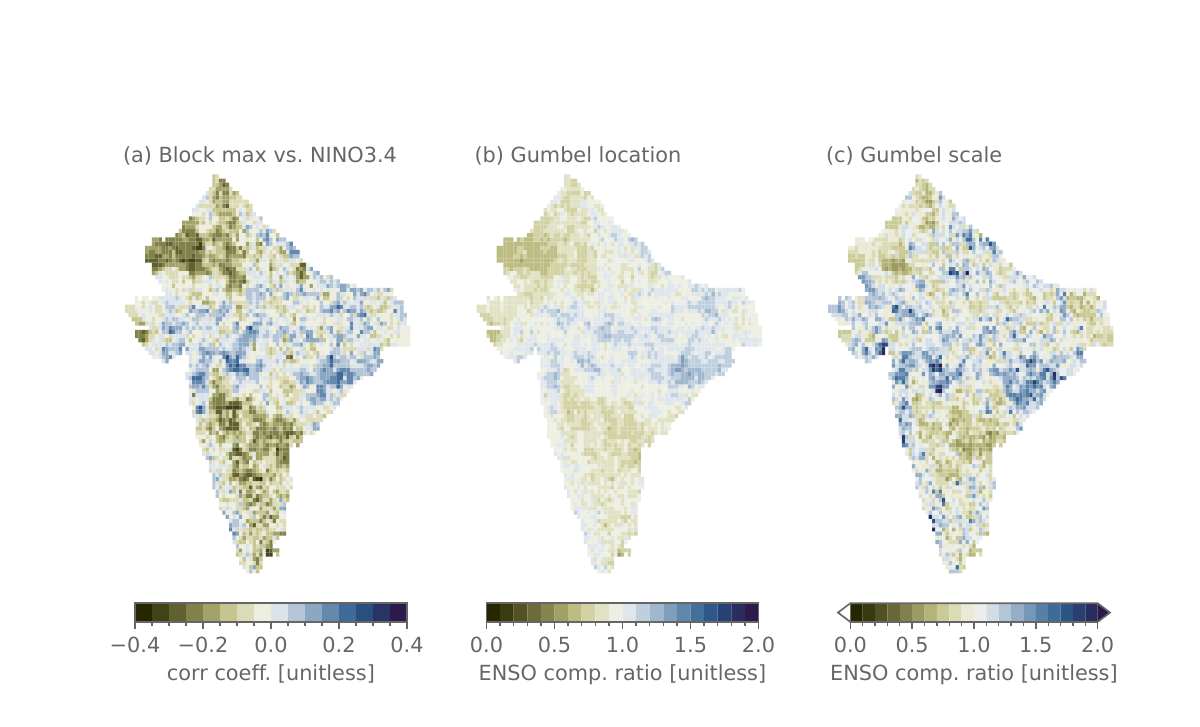}\\
\includegraphics[width=0.66\textwidth]{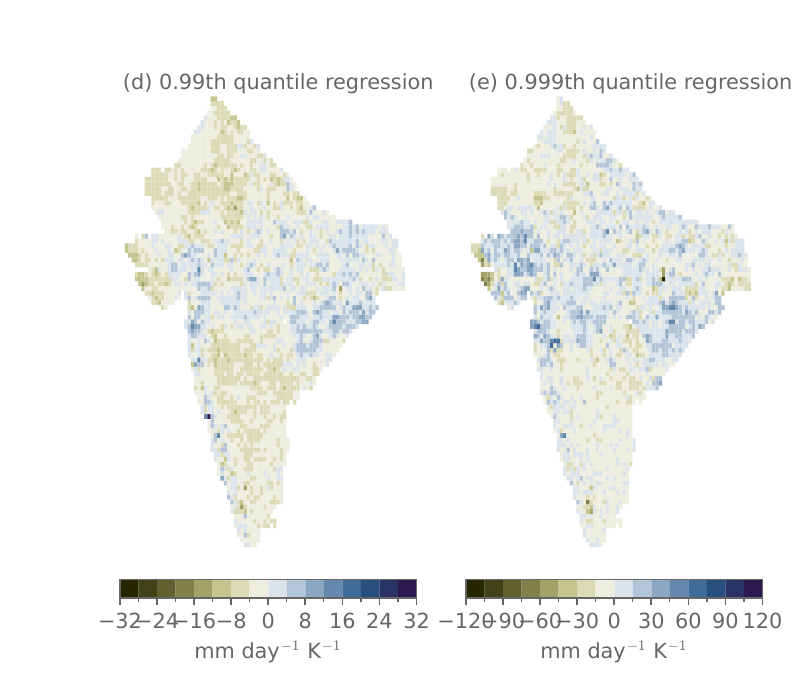}
\caption{(a) Pearson correlation coefficient between the JJAS block maximum daily rainfall and NINO3.4, 1901-2020.  (b and c) Ratio of \elnino/ to \lanina/ composite Gumbel fits to JJAS block maximum daily rainfall at each gridpoint, for the (b) location and (c) scale parameters.  (d and e) Quantile regression for the (c) 99th and (d) 99.9th percentile daily rainfall (including both rainy and dry days) against NINO3.4.}
\label{fig:sens-tests}
\end{figure*}

\newpage
\begin{figure*}
\centering
\adjustimage{valign=c,width=0.6\textwidth}{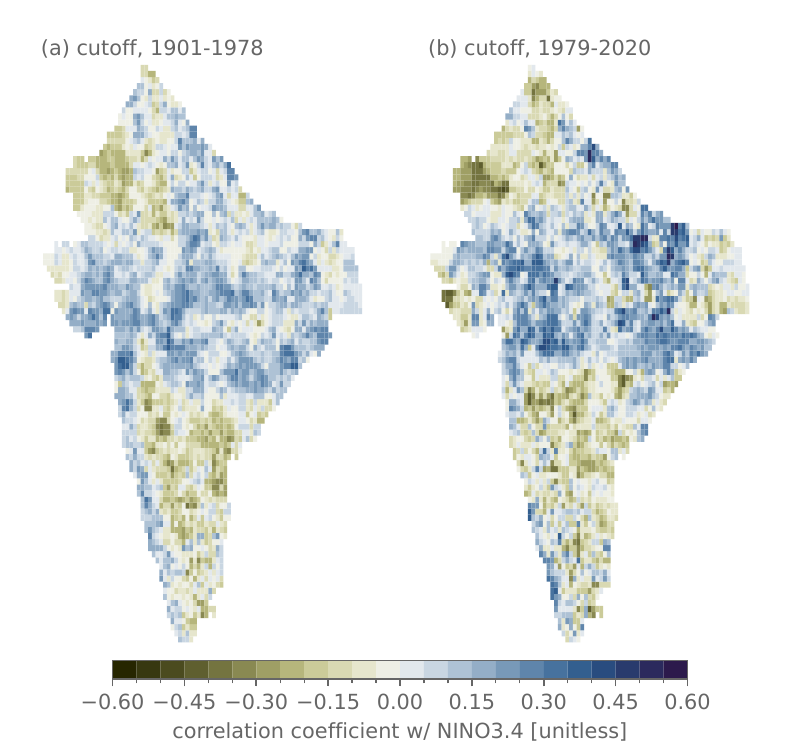}
\adjustimage{valign=c,width=0.39\textwidth}{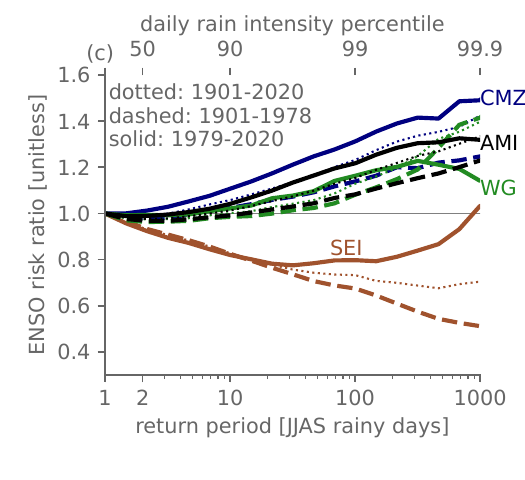}\\
\includegraphics[width=0.7\textwidth]{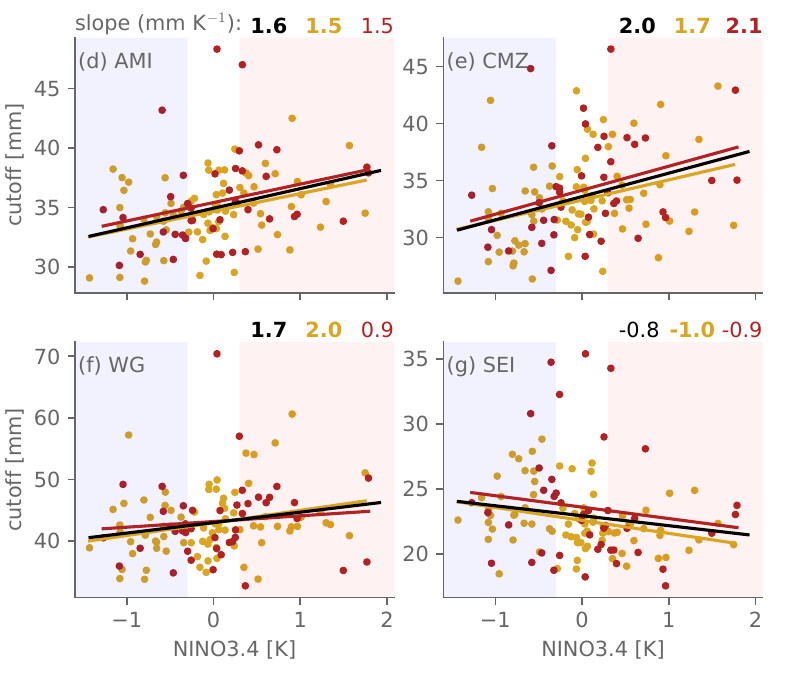}\\
\caption{(a and b) Pointwise correlations between the JJAS cutoff and NINO3.4 restricting to (a) the pre-satellite era of 1901-1978 and (b) the satellite era of 1979-2020.  (c) ENSO risk ratios for each region restricting to (dashed) the pre-satellite era and (solid) the satellite era.  (d through g) Scatterplots of the region-aggregated cutoff in each JJAS \vs/ NINO3.4, for (d) All Monsoonal India, (e) Central Monsoon Zone, (f) Western Ghats, and (g) Southeast India.  Overlaid lines are corresponding linear regressions.  Yellow elements are for the pre-satellite era, dark red elements are for the satellite era, and black elements are for the full period.  Printed values above each panel are the regression slopes for that region, in mm K$^{-1}$, with values in bold statistically significant at the \(p=0.05\) level.  Light red and blue shading denote \elnino/ and \lanina/ conditions based on the \(\pm0.3\)~K threshold used elsewhere for constructing composites.  Note differing vertical axis spans across d-f.}
\label{fig:presat-sat}
\end{figure*}

\begin{figure*}
\centering
\includegraphics[width=\textwidth]{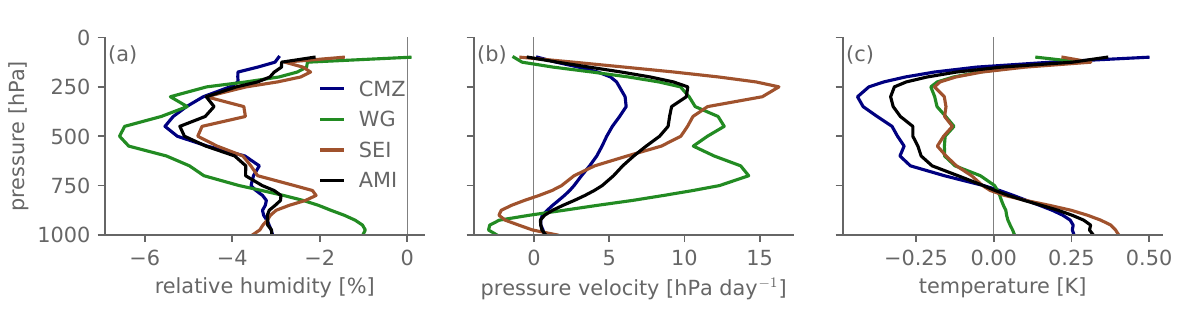}
\caption{JJAS-mean ENSO composite anomaly profiles, \elnino/ minus \lanina/, averaged over the AMI domain, in relative humidity and vertical velocity.}
\label{fig:enso-mean}
\end{figure*}

\begin{figure*}
\centering
\includegraphics[width=\textwidth]{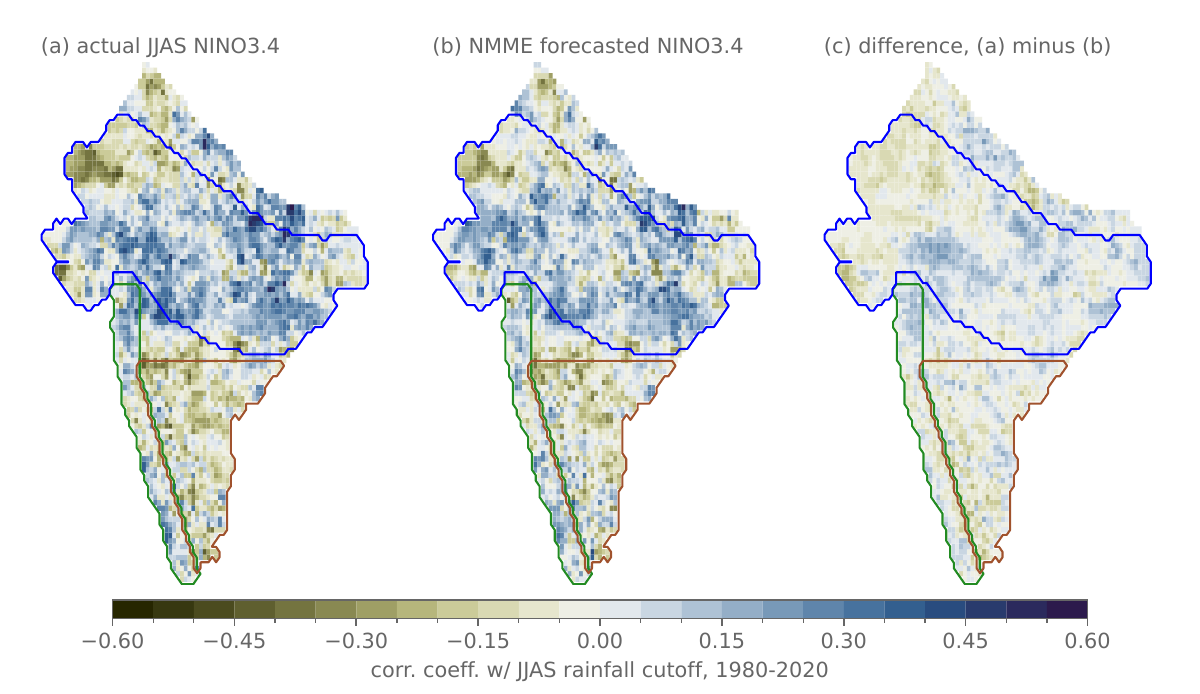}
\caption{For the 1980-2020 period for which NMME data is available, correlation coefficient of the local JJAS cutoff \vs/ the JJAS NINO3.4 value computed (a) directly from ERSST, or (b) forecasted by the NMME multi-model mean of runs initialized at the preceding May 1st.  Panel (c) shows their difference.}
\label{fig:nmme}
\end{figure*}

\end{document}